# Spike conformation transition in SARS-CoV-2 infection

Liaofu Luo* (College of Physical Sciences, Inner Mongolia University, Hohhot, China)

Yongchun Zuo (College of Life Sciences, Inner Mongolia University, Hohhot, China)

* Correspondence address:   lolfcm @ imu.edu.cn

**Abstract**   A theory on the conformation transition for SARS-CoV-2 spike protein (S) is proposed. The conformation equilibrium between open (up) and closed (down) conformations of receptor binding domain (RBD) of the spike is studied from the first-principle. The conformational state population is deduced from the free energy change in conformation transition of S protein. We demonstrated that the free energy includes two parts, one from the multi-minima of conformational potential and another from the variation of structural elasticity. Both factors are dependent of amino acid mutation. The former is related to the change of affinity of RBD to ACE 2 due to the mutation in the subdomain RBM (receptor binding motif) of RBD. The latter is caused by the change of elastic energy of S protein. When the affinity has increased significantly and/or the elastic energy has been reduced substantially the equilibrium is biased to the open conformation. Only then can the virus infection process continue. Possible new SARS-CoV-2 variants from amino acid mutations in 5-9 sites on RBD interface are predicted. The elastic energy variation needed for conformation transition is estimated quantitatively. Taking the elastic-structural change into account more virus variants are possible. Both the D614G mutation, the K986P mutation and the new variants 501Y in current SARS-CoV-2 pandemic can be interpreted from the presented theory. The comparison of the infectivity of SARS-CoV-2 with SARS-CoV-1 is made from the point of conformation equilibrium. Why the virus entrance takes priority in lower temperature and higher humidity is interpreted by the present theory. The conformational transition influenced by electromagnetic field is also discussed briefly.

## 1   Introduction

The current SARS-CoV-2 pandemic has created urgent needs for diagnostics, therapeutics and vaccines. However, meeting these needs firstly requires a deep understanding of the mechanism of viral infection. For all enveloped viruses, membrane fusion is a key early step for entering host cells and establishing infection. The surface of coronaviruses are decorated with a spike (S) protein (about 1273 amino acids for SARS-CoV-2), a large class 1 fusion protein. The S protein forms a trimeric complex that can be functionally categorized into two distinct subunits S1 and S2 subunits . There is a receptor binding domain (RBD) in S1 (between sequence sites 331 to 528 for SARS-CoV-2) which interacts with a host-cell receptor protein to trigger membrane fusion. The host-cell receptors for SARS-CoV and MERS-CoV are angiotensin-converting enzyme 2 (ACE2) and DPP4 respectively [1][2]. Recent advances in cryo-electron microscopy (cryo-EM) characterization of the spike protein revealed that the RBDs adopted at least two distinct conformations. RBD can be either in the open or in the closed position (called up or down conformation respectively).[3] In the up conformation, the RBD jut out away from the rest of S, such that they can easily bind with ACE2. In the down conformation the RBDs are tightly packed, preventing binding by ACE2.[4] The receptor-binding event may trap the RBD in the less-stable up conformation, leading to destabilization of S1,triggering the conformational change of S2 from prefusion to postfusion state and initiating the membrane fusion. The SARS-CoV cell entry also depends on transmembrane protease serine 2 (TMPRSS2) etc which help to cut S to units S1 and S2.[5].

The above conformational transition processes can be expressed through a set of equations

$$RBD(closed) \longleftrightarrow RBD(open) \qquad (1)$$

$$RBD(open) + ACE2 \longrightarrow AS \qquad (2)$$

$$S + E \rightarrow S_1 + S_2 + E$$

$$S2(prefus) \longrightarrow S2(postfus) \qquad (3)$$

where AS in Eq (2) denotes the RBD-receptor bound state. Eq (3) is a set of equations causing the transition of S protein from prefusion to postfusion state and in its first equation E denotes the enzyme TMPRSS2 . The di-directional transition of Eq (1) makes RBD attaining at an equilibrium between open and closed conformations. Eq (1) is the starting point of the viral infection, we shall focus on it in the paper.

## 2  Method - Mathematical relation of conformation equilibrium

Eq(1)describes the conformation transition of RBD. The closed/open transition of RBD proceeds in two directions the equilibrium of which determines the first step of viral infection. We shall study the conformation equilibrium between RBD(closed) and RBD(open).  Suppose A denotes closed conformation and B open conformation. The Gibbs free energy is expressed by $G_A$ and $G_B$ respectively in two conformations.  Generally, if $G_A$ is lower than the $G_B$, then S1 takes inactive conformation A with large probability. As the spike protein located in this structure the subsequent steps of Eqs (2) and (3) will not start and the virus infectious process cannot continue. Oppositely, in order to start Eq(2), the equilibrium of Eq(1) should be biased to the right open conformation, that is, $G_B$ should be lower than $G_A$.

The free energy G in a given conformation is related to the partition function Z

$$G = -k_B T \ln Z$$
$$Z = \sum_n e^{-\beta E_n} \quad (\beta = \frac{1}{k_B T}) \tag{4}$$

where $E_n$ is the energy level in the conformation and the sum in Eq(4) is over the quantum states $n$.[6] Thus we have the free energy increase $\Delta G = (G_B - G_A)$ as

$$\Delta G = -k_B T \ln \frac{Z_B}{Z_A} \tag{5}$$

$\Delta G$ determines the conformational state population. $\Delta G > 0$ means conformation A more populated and $\Delta G < 0$ B more populated. Set $E_n$ for the molecule in conformation A or B is denoted by $(E_n)_A$ or $(E_n)_B$ respectively,

$$(E_n)_{A,B} = V_{A,B} + (n + \frac{1}{2})\hbar \omega_{A,B}. \tag{6}$$

where $V_{A,B}$ is the conformational energy for the molecule in state A or B which is determined by two minima of conformational potential respectively and $(n+\frac{1}{2})\hbar\omega_{A,B}$ the corresponding vibrational energy around the minimum of conformational potential (called elastic energy hereafter). By the summation of Boltzmann factor over vibration states one has [7]

$$\frac{Z_A}{Z_B} = e^{-\beta(V_A - V_B)} Y_{A/B} \quad (\beta = \frac{1}{k_B T}) \tag{7}$$

$$Y_{A/B} = \frac{e^{\frac{1}{2}\beta\hbar\omega_B} - e^{-\frac{1}{2}\beta\hbar\omega_B}}{e^{\frac{1}{2}\beta\hbar\omega_A} - e^{-\frac{1}{2}\beta\hbar\omega_A}} \tag{8}$$

Since the partition function is derived from the summation of Boltzmann factor, $Z_A / Z_B$ given by Eq(7) means the conformational state population, the ratio of the probabilities in two conformations A and B. It shows that if the vibration is neglected then the population is simply determined by the symbol of $V_A - V_B$. For example, A is the advantageous conformation as

$V_A < V_B$. However, equation (7) shows that the vibrations around the minimum of potential are important for the conformational state population. In fact, from Eq (8) one has

$$Y_{A/B} = \frac{\omega_B}{\omega_A} \quad \text{as} \quad k_B T \gg \hbar\omega_A, \hbar\omega_B \tag{8.1}$$

$$Y_{A/B} = \exp\{\frac{\hbar(\omega_B - \omega_A)}{2k_B T}\} \quad \text{as} \quad k_B T \ll \hbar\omega_A, \hbar\omega_B \tag{8.2}$$

In general

$Y_{A/B} > 1$ and decreasing monotonously with $T$ as $\omega_A < \omega_B$

$0 < Y_{A/B} < 1$ and increasing monotonously with $T$ as $\omega_A > \omega_B$ (9)

Inserting (8) into (7) we have

$$\Delta G = E_{conf} + E_{elas}$$

$$E_{conf} = V_B - V_A$$

$$E_{elas} = k_B T \ln Y_{A/B} \tag{10}$$

The right-hand-side of Eq(10) contains two parts: $E_{conf}$ describes the conformational energy, and ω-related term $E_{elas}$ describes the elastic energy. Thus by taking the elastic energy into account the conformation with lower vibration frequency increases its probability. When $\omega_B$ is much smaller than $\omega_A$, in spite of $V_A < V_B$ one always has $Z_A/Z_B < 1$ and the conformation B is the advantageous (with large probability) instead of A. (see sketch map of conformational potential, Fig 1)

Another relation can be used in estimating the elastic energy term $Y_{A/B}$ is

$$k_B T \ln Y_{A/B} = \frac{\hbar(\omega_B - \omega_A)}{2} \text{ctnh} \frac{\hbar\omega_A}{2k_B T} \quad (as \quad \left|\frac{\omega_A - \omega_B}{\omega_A}\right| \ll 1) \tag{11}$$

where the function ctnh$x$ is defined by $\text{ctnh}x = \frac{e^x + e^{-x}}{e^x - e^{-x}}$, an odd function decreasing with $x$ and always larger than 1 for positive $x$.

Consider amino acid mutations in spike. The single amino acid mutation causes changes both in conformational energy $E_{comf}$ and elastic energy $E_{elas}$. There are two ways to make conformation transition from closed to open: one is decrease of conformational energy $E_{comf}$, and another is reduce of elastic energy $E_{elas}$. As $\Delta G < 0$ we have $(Z_B/Z_A) > 0$ and the equilibrium between two conformations shifts to the side of conformation B; otherwise, as $\Delta G > 0$ the equilibrium remains tend to the conformation A.

## 3  Results

The conformational potential and structural elasticity of spike protein is dependent of its amino acid sequence. One may increase or decrease the conformational and elastic energy through amino acid mutation. Therefore, the originally inactive conformation "up" can transform into an active advantageous "down" by selecting appropriate amino acid mutation. Recently the deep mutational scanning of SARS-CoV-2 RBD was completed and its constraints on folding and ACE2 binding were analyzed [8]. However, how the mutation influences the conformation equilibrium and in turn, influences the virus infection, has not been studied. In the following we will sketch the main results of our method for explaining experimental facts.

### 3.1) Amino acid mutations at RBD ACE2 interface

The transition from closed to open conformation can occur by decreasing conformational energy $E_{conf} = V_B - V_A$ through amino acid mutation. It can be attained through the mutational change of affinity for about 20 residues at RBD ACE 2 interface in the subdomain RBM (receptor binding motif) (see Table 1). The affinity given in the table is the value relative to unmutated variant. We found for most mutational sites at interface the affinity change takes a negative value. However, they take a large positive affinity at several sites that predicts a possible conformational transition. We predict the amino acid mutations with higher affinity and higher expression as new variants of SARS-CoV-2 as shown in the table. We found that there are nine sites where the amino acid mutation possibly leads to the conformation transition of SARS-CoV-2 and five of which, namely at site 501,498,493,503 and 484 respectively require special attention. The predicted mutation N501Y has been observed in the variant 20B/501Y.V1(or B.1.1.7) in U.K., the variant 20C/501Y.V2(or B.1.351) in South Africa and the lineage B.1.1.248 in Brazil. The predicted mutation E484K has been observed in the variant 20C/501Y.V2 in South Africa. [9]

**Table 1  Affinity change at RBD ACE2 interface
in amino acid mutation**

| mutational site | K 417 | N 439 | G 446 | Y 449 | Y 453 | L 455 | F 456 | A 475 | E 484 | F 486 | N 487 |
|---|---|---|---|---|---|---|---|---|---|---|---|
| affinity sum | none | 0.04 | none | none | 0.33 | 0.05 | none | none | 0.29 | none | none |
| pred variants | | | | | Y453F | | | | E484R E484K | | |
| SARS-CoV1 | V | R | T | Y | Y | Y | L | P | P | L | N |
| affinity | -0.32 | -0.03 | -0.19 | 0 | 0 | -1.5 | -0.11 | -1.62 | -0.28 | -0.47 | 0 |

| mutationai site | Y 489 | Q 493 | S 494 | G 496 | Q 498 | T 500 | N 501 | G 502 | V 503 | Y 505 | L 517 |
|---|---|---|---|---|---|---|---|---|---|---|---|
| affinity sum | none | 0.65 | 0.19 | none | 0.68 | none | 0.89 | none | 0.29 | 0.13 | 0.23 |
| pred variants | | Q493M Q493A | S494H | | Q498H Q498Y Q498F | | N501F N501Y N501V N501W | | V503K V503M | Y505W | L517M |
| SARS-CoV1 | Y | N | D | G | Y | T | T | G | I | Y | L |
| affinity | 0 | -0.21 | -1.1 | 0 | 0.16 | 0 | 0.1 | 0 | 0.05 | 0 | 0 |

The amino acid sites at ACE2 interface are given in the first line which are taken from [8]. Note that 484E and 517L are in the position adjacent to the interface. The affinity to ACE 2 is represented by $\Delta \log K_p$ relative to the unmutated SARS-CoV-2 RBD where $K_p$ is dissociation constant [8]. The sum of all positive affinities for amino acid mutation on the site is given in the second line and "none" means none mutation with positive affinity. The predicted variants in the third line are selected according to the conditions $\Delta \log K_p > 0.05$ and expression $\Delta \log MFI > -0.2$ where MFI- mean fluorescence intensity [8]. The amino acid replacement in SARS-CoV-1 is given in the fourth line and the affinity for that replacement is given in the fifth line .

### 3.2)  Amino acid mutations outside RBM subdomain

**D614G mutation**   In the following three sections we will adduce evidence to show that the conformation transition occurs through the approach of elastic energy change. There is strong evidence that the spike protein mutation D614G increases infectivity of the COVID-19 virus and the G614 variant has become the most prevalent form in the global pandemic [10]. How to explain this important event in the virus evolution? The recent analysis on the structure and function of the D614G spike protein indicated that although the D614G affinity for ACE2 is reduced the conformation can still be shifted towards an ACE2 binding-competent state[11]. This is consistent with our conformation equilibrium theory mentioned above, since the mutation D614G removes the H-bond between residues D614 in S1 and T859 in S2 and the RBD structural elasticity has been changed and the flexibility has been increased. For the wild type spike the RBD is in closed conformation A, where $V_A$ is lower than $V_B$. However, the elasticity decreases largely due to amino acid replacement and the conformation B changes to the advantageous one. It means the amino acid mutation triggers the transition of RBD from a closed to a more open conformation. The open RBD will be able to initiate the subsequent processes following Eq (2) and Eq(3) and consequently the spike protein changes itself from prefusion to postfusion states to mediate fusion of viral and cellular membrane. This interprets why the mutation D614G largely increases the infectivity of the COVID-19 virus.

**K986 P mutation** In recent mRNA vaccine design people demonstrated that mRNA expressing SARS-CoV2S-2P is a potent immunogen. They identified 2 proline substitutions (2P) at the apex of the central helix and heptal repeat 1 (HR1) that effectively stabilized MERS-CoV and SARS-CoV proteins in the prefusion conformation [12]. The prefusion-stabilized protein immunogens that preserve neutralization-sensitive epitopes are an effective vaccine strategy for enveloped viruses. Here the key step is the 2P mutation K986 P and V987P in SARS-CoV2 spike sequence. The K986 P mutation removes a salt bridge between Lys986 and either ASP427 or ASP428 of another protomer in the trimer interface [2]. The mutation relaxes the structure of S protein. That is, due to the mutation the structural elasticity of the spike decreases and in turn, it jumps from the closed conformation A to the open conformation B. Since B is an easily infectious conformation, the K986 P mutation provides an immunogen design that can trigger immediate rapid manufacturing of an mRNA vaccine.

**Several commonly studied amino acid variants related to spike elastic energy** Above two examples show that, apart from conformational-energy-related mutations in RBM subdomain the amino acid mutations in other spike regions also trigger the conformation transition. These mutations are related to the elasticity change of the spike. The spike elastic energy plays important role in studying SARS-CoV-2 infection. Several commonly studied amino acid mutations in the spike regions outside RBM subdomain are listed in Table 2. The global frequencies of these amino acid variants are given in the table and the corresponding free energy change due to elastic-energy are also estimated and listed in the table.

**Table 2  Amino acid variants related to spike elastic energy**

| Mutation number | Spike mutation | Region | Count | $-\Delta G/k_B T$ |
|---|---|---|---|---|
| 1 | D614G | S1 CTD domain | 71% | 0.895 |
| 2 | L5F | Signal peptide | 0.6% | -5.11 |
| 3 | R21I/K/T | S1 NTD domain | 0.5% | -5.29 |
| 4 | A829T/S | Fusion peptide | 0.3% | -5.81 |
| 5 | D839Y/N/E | Fusion peptide | 0.5% | -5.29 |
| 6 | D936Y/H | HR1 | 0.9% | -4.70 |
| 7 | P1263L | Cytoplasmic tail | 0.7% | -4.95 |
| 8 | K986P | HR1 | >70% | >0.85 |
| 9 | P681H | S1 near cleavage | large | positive |
| 10 | 69-70 del | S1 NTD domain | large | positive |

Experimental data of mutations 1-7 are taken from [10]. Data of mutation 8 are taken from [2][12]. Data of mutations 9 and 10 are taken from [9]. $-\Delta G/k_B T$ in the last column is calculated by using Eq(5) where $Z_A/Z_B$ is taken from the experimental count of population. The lower bound of count in mutations 8 is estimated from thermostable spike trimer where about 80% of S-R/PP trimers have one open RBD [13]. The mutations 9 and 10 are associated with N501Y in the variant B.1.1.7 where the separate effect of elastic energy has not been defined quantitatively (section 3.3).

From Table 2 we found the free energy increase $\Delta G$ of open conformation relative to closed is positive in most mutations apart from D614G, K986P, P681H and 69-70 del. It explains why a substantial number of mutations have not been selected in current SARS-CoV-2 pandemic isolates. Moreover, Table 2 indicates $-\Delta G/k_B T$ always takes a positive value near +1 or takes a negative value near -5 for all mutations. The large negative value means the amino acid mutation having not brought any obvious change of elastic energy so that $\Delta G$ is dominated by $(V_B-V_A)$ term and remains positive. While the former case of positive value means the amino acid mutation makes the elastic energy of open conformation much lower than the closed conformation so that $\Delta G$ changes to negative.

How to explain the relation between the amino acid mutation and the change of spike elasticity?  We know that 20 amino acid residues are classified into three categories : the acid residues (as anions) D,E and Y; the basic residues (as cations) H,K and R; and others – neutral residues. One may assume a significant change of elastic energy can be found only if the residue's electric charge difference in the mutation is 1 unit or more and otherwise, the elastic energy variation is too small so that the mutation cannot trigger the structural changes of S protein. From the experimental data in Tab 2 we know that in the former case $-\Delta G/k_B T$ takes a positive value near +1 and in the latter case $-\Delta G/k_B T$ takes a negative value near -5. Therefore the electricity change in the mutation determines the elasticity change of the spike and finally determine which one (open or closed) is the advantageous conformation of RBD.

### 3.3) New variant strains 501Y

The new variant 20B/501Y.V1 (or B.1.1.7) emerged in the fall of 2020 [9]. Apart from D614G, it includes the mutation N501Y in RBD, the spike deletion 69-70del and the mutation P681H adjacent to the furin cleavage site. The mutation N501Y contributes a large positive value of affinity and beneficial to conformation transition from closed to open. The deletion 69-70 in association with other RBD changes may relax the spike structure and decrease the elastic energy. The mutation P681H is immediately adjacent to the furin cleavage site which may help the S protein cut to S1 and S2 under the enzyme action. All these factors are superimposed together and make the new variant strain emerging with an unusually high growth rate. The variant was firstly detected in the U.K. with high infectivity and rapidly detected in numerous countries around the world. Later, another variant 20C/501Y.V2 in South Africa and another lineage B.1.1.248 in Brazil were reported independently. Both they include the mutation N501Y in RBD. Besides, the mutant E484K was detected in 20C/501Y.V2 which has a strong ability to escape from neutralizing antibodies that may be related to the gene network of immune escape [13].

### 3.4) More conformations of virus S protein

Three conformations of the prefusion trimer observed by using cryo-electron microscopy on intact virions: all RBDs in the closed position ; one RBD in the open position ; and two RBDs in the open position.[3] However, the two-open conformation has only been observed in vitro after inserting multiple stabilizing mutations. Moreover, through the designed mutations in S protein these authors observed distinct closed and locked conformations of the S trimer and the classification of the cryo-EM data showed that the disulfide bond formation is beneficial to the closed RBDs.[14] The above experimental data can be explained as well by our model. Disulfide bond is a strong bond. The disulfide bond formation gives another factor to tighten the S protein structure in addition to commonly the H-bond and salt bridge. Following the present model the multiple stabilizing mutations can be explained by the conformational potential $U(\theta)$ of three or more minima. The generalized multi-minima can describe the abundant structures of S protein and many new conformations of RBD appeared in the experiment. Any change of the vibration frequency ω around one minimum will give additional contribution to the free energy of the prefusion trimer. The conformation equilibrium among multi-minima will provide a starting point for the study of the phase-transition dynamics of the system.

## 4  Discussions

4.1）*Conformational equilibrium of spike protein in amino acid mutation*. The conformational change of the spike protein is of special significance for virus infection. To understand the virus entry and viral infection one needs know how the conformational equilibrium of the spike protein takes place and how the amino acid mutation changes the equilibrium. We found that the SARS-CoV-2 infection of humans and its pandemic is always through the amino acid mutation on the spike. One approach is the binding affinity of RBD to ACE 2 increasing due to the amino acid mutation in the subdomain RBM. Another approach is the elastic energy reduction due to the change of the structure, for example, the loss of some hydrogen bonds or salt bonds of the spike protein. The former is related to the change of conformational potential energy $V_B-V_A$, while the latter is related to the variation of conformational vibration frequency difference $\omega_A-\omega_B$ or frequency ratio $\omega_A/\omega_B$. We found there exist several amino acid mutations on RBD interface that may produce the new SARS-CoV-2 variants (see Table 1). The elastic energy variation needed for conformation transition can be estimated as follows. From Table 2 we found that, as the amino acid mutation does not cause the breaking of hydrogen bond or salt bond, the free energy difference $\Delta G$ between closed and open conformations is about $-5k_BT$, while as the amino acid mutation causes the break of hydrogen bond or salt bond, $\Delta G$ is near $+1\ k_BT$. The difference $6k_BT$ between two cases should come from the elastic energy variation. Therefore, the hydrogen bond (salt bond) break in spike protein generally needs energy input about $6k_BT \backsim 0.15$ ev which gives reasonable order-of-magnitude estimate.

4.2) *Comparison of conformation equilibrium in SARS-CoV-2 and in SARS-CoV-1*. Why does SARS-CoV-2 infect humans so easily? Deep mutational scanning of all amino acid mutations gives new understanding on the problem, suggesting that there is a substantial mutational space consistent with sufficient affinity to maintain human infectivity [8]. RBD ACE2 interface is a region where amino acids are prone to mutation. The evolutionary conservatism between two interfaces of SARS-CoV-2 and SARS-CoV-1 is 40%, lower than the conservatism 79% for the total genome comparison. The affinity comparison between SARS-CoV-2 and SARS-CoV-1 are given in Table 1. There are 12 mutational sites in total in the subdomain RBM: three sites at 498, 501 and 503 showing nearly equal affinity and other nine sites of SARS-CoV-2 having higher affinity than SARS-CoV-1. It means that on average, SARS-CoV-2 RBD binds ACE2 stronger than SARS-CoV-1. This is the first reason why the infectivity of SARS-CoV-2 is higher than SARS-CoV-1. When we study how the amino acid mutation influences the virus infectivity we should consider the contribution from structural elasticity variation in the total RBD. Comparing RBD sequences between SARS-CoV-1 and SARS-CoV-2 we found there are 49 single amino acid mutations in total and fourteen of them have the electric charge variation of 1 unit. These mutations indicate that the variation of structural elasticity of the spike does exist in the evolution of SARS-CoV-1 and SARS-CoV-2. Moreover, we found eight mutations from SARS-CoV-1 to SARS-CoV-2 (namely E354N, R439N, K452L, Y455L, D476G, K478T, D494S and Y498Q) are of charge-decreasing type (from charge negative or positive to neutral), and six mutations (V417K, N460K,V471E, F473Y, P484E and N519H) are of electric charge-increasing type (from charge neutrality to negative or positive).  It can be roughly estimated that the elasticity-decreasing mutations occurred about 14% more in SARS CoV-2 than in SARS CoV-1 evolution.   This is the second reason to explain the infectivity of SARS-CoV-2 higher than SARS-CoV-1 from the point of conformation equilibrium.

4.3）*Conformational transition induced by electromagnetic field* . As shown in Fig 1 there is a potential barrier between two conformations. The transmission coefficient is dependent of the height and width of the potential barrier. Introducing an electric field will change the conformational barrier of the spike effectively and in turn change the conformational state population. It was reported that as the electric field increases beyond 0.02 au, the net electron density starts to move from C-H bond towards the carbon, causing the bond to begin to weaken and lengthen. Thus, the static electric field of appropriate strength and direction can break some H-bond and salt bond in the spike and changes the conformation equilibrium of RBD. On the other hand, the geomagnetic field can effectively block the bombardment of high-energy charged particles in the cosmic ray. Due to abnormal macula activity the weakened geomagnetic field would make the spike susceptible to bombardment, causing the residue deletion or amino acid mutation and changing the conformational state population.

4.4）*Equilibrium between RBD closed and open conformations correlated with temperature and humidity*. Conformational equilibrium of RBD is dependent of temperature, that can be seen from Eqs (7) and (8). From Eq (9) we know that $Y_{A/B}$ depends on temperature monotonously, decreasing with $T$ as $\omega_A<\omega_B$ and increasing with $T$ as $\omega_A>\omega_B$) . Define the phase transition temperature $Tc$. $Tc$ is determined by $\Delta G=0$. From Eqs (10) (11) we obtain a simplified equation for $Tc$

$$\frac{2(V_B - V_A)}{\hbar(\omega_A - \omega_B)} = ctnh \frac{\hbar\omega_A}{2k_B T_c} \qquad (12)$$

Eq (12) gives a curve on $\frac{2(V_B - V_A)}{\hbar(\omega_A - \omega_B)}$ (y) - $\frac{\hbar\omega_A}{2k_B T}$ (x) plane that divides the plane into two phases, closed conformation and open conformation. Under given parameters $V_A$, $V_B$ , $\omega_A$, and $\omega_B$ , when the environmental temperature decreases to $T<Tc$ the conformational transition from closed conformation to open occurs immediately. This explains why the virus entrance takes priority in the winter. Oppositely, the best time to eliminate the virus is in summer. Apart from temperature, the conformational equilibrium depends on humidity. It was reported that the soluble S trimer with the PP mutation has a looser structure than the full-length S with wild-type sequence [2] . So, the moisture may be conductive to K986 P mutation and makes the virus infectious. In fact, the virus can be modeled as a charged sphere. From the electrostatics for salty solution, one can arrive at an expression for the potential at the surface of the charged sphere and the dielectric constant $\epsilon$ has entered into the expression of the potential [15].   It means the elastic frequency $\omega^2$ should be replaced by $\omega^2/\epsilon$. For water $\epsilon=80$. Therefore, the frequency parameter takes a reduced value $\omega/9$ in the full salty solution instead of $\omega$ in vacuum. In this way we estimate the elastic frequency should be reduced by a multiple up to 9 in the humid environment. This gives a quantitative estimate on the virus infection strongly dependent of humidity.

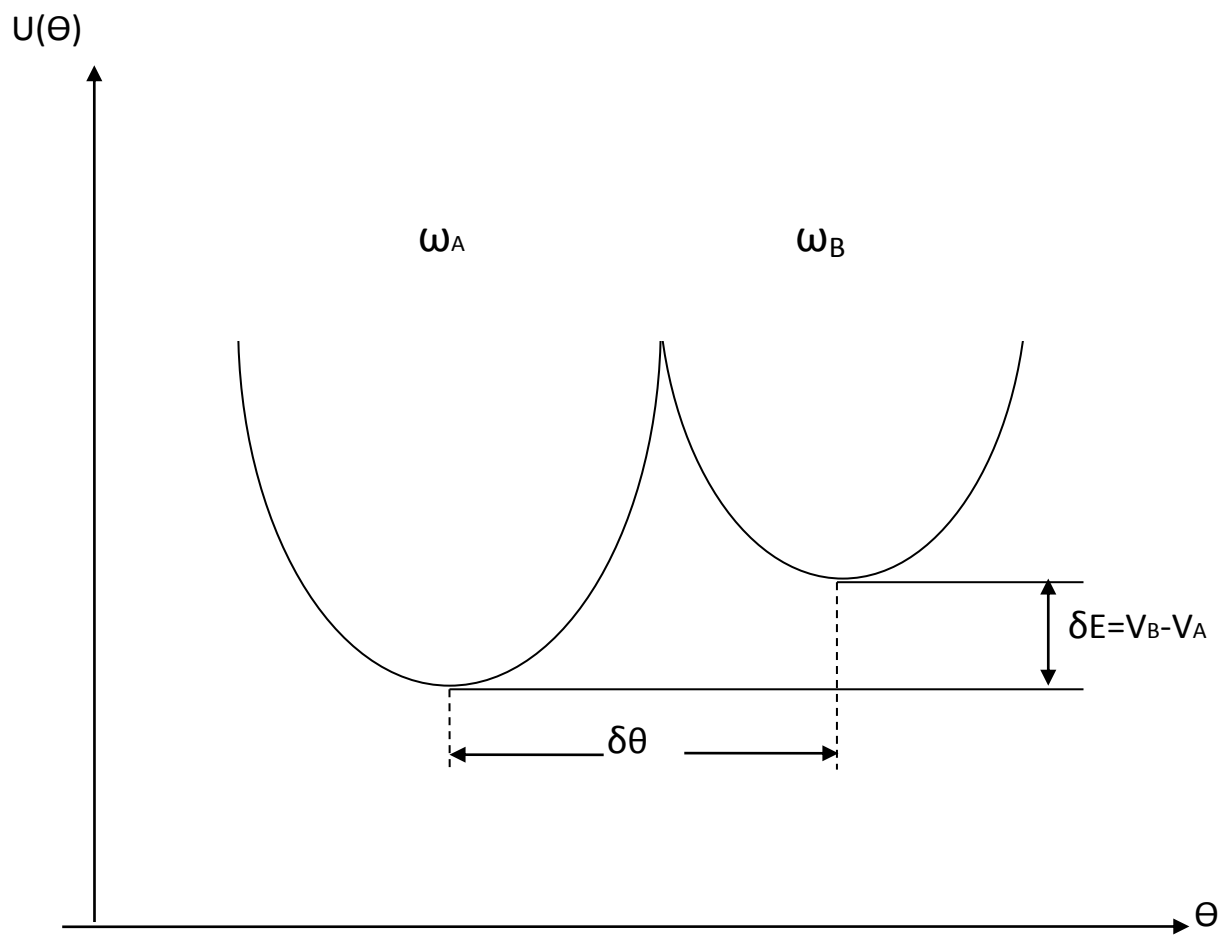

**Fig 1  Sketch map for conformational Potential U(θ)**

$U(\theta) = V_A + \dfrac{1}{2} I\omega_A^2 (\theta - \theta_A)^2$ (left), $\qquad U(\theta) = V_B + \dfrac{1}{2} I\omega_B^2 (\theta - \theta_B)^2$ (right)

Two minima separated by a potential barrier represent two conformations A (left) and B (right) of RBD, $\theta$ is conformational coordinate, $\omega$ – the frequency parameter and $I$ – the inertia parameter describing the vibration around the minimum. The molecule may located in conformation A or B. The conformational state population is determined by two factors: the conformational energy $E_{conf} = V_B - V_A$ and the elastic energy $E_{elas} = k_B T \ln Y_{A/B}$. The elastic energy is related to the frequency-ratio $\omega_B/\omega_A$ or frequency-difference $\omega_B - \omega_A$. As $V_A < V_B$, A is the advantageous conformation if the elastic energy can be neglected (i.e. $\omega_B \cong \omega_A$). However, when $\omega_B$ is much smaller than $\omega_A$ the contribution of elastic energy is important and it can make the population conversion, the conformational state B being advantageous.